\documentclass[11pt]{amsart}

\usepackage{amsmath,amsthm,amsfonts,amssymb}
\usepackage{graphicx}
\usepackage{hyperref}
\usepackage[usenames]{color}

\theoremstyle{definition}

\newcounter{comcount}

\numberwithin{equation}{section}

\title{
Cryptanalysis of two schemes of Baba et al. by linear algebra methods }

\author{Vitali\u{i} Roman'kov}

\address{Dostoevsky Omsk State University}
\email{romankov48@mail.ru}

\date{}

\begin{document}

\maketitle

\footnote{Supported by RFBR, project 18-41-550001.}

\begin{abstract}
We show that the attacks based on the linear decomposition method introduced by the  author  and the span-method introduced by Tsaban  allow one to find  the transmitted message   in the cryptosystem and  the exchanged key in the protocol which are  proposed in \cite{BKT}. 
\end{abstract}

\section{Introduction}

 In \cite{BKT}, S. Baba, S. Kotyada and R. Teja  demonstrate how to define an approximate  one-way function FACTOR in a non-Abelian group. As examples of a platform for realization of FACTOR they suggest  one of groups like GL$_n$($\mathbb{F}_q$), UT$_n$($\mathbb{F}_q$), or Braid Groups $B_n$, $n \in \mathbb{N}.$ Here $\mathbb{F}_q$ denotes the finite field of order $q.$  They believe that the function FACTOR is one-way. It means  that the inverse to the FACTOR is easy to compute, while the function itself is hard to compute. 
 
  Then, using FACTOR function as a primitive the authors of \cite{BKT} therefore define a public key cryptosystem which is comparable to the classical El-Gamal system  based on the discrete logarithm problem. Recall, that the El-Gamal system can be described as follows: Let $G$ be a public finite cyclic group with generator $g$, and let  $x \in \mathbb{Z}$  is Alice’s private key. The element $g^x$ is public. To send a message $m \in  G,$ Bob picks a random   integer $y$ and sends the cipher text $c = (g^y,g^{xy}m)$ to Alice. To decrypt, Alice calculates $(g^y)^x=g^{xy}$ and inverts it to retrieve $m.$ 
  
  In \cite{BKT}, the authors also propose a key exchange, analagous to the Diffie-Hellman key exchange protocol  in a non-Abelian setting using FACTOR. Recall, that the classical Diffie-Hellman protocol  can be described as follows: Let $G$ be a public finite cyclic group with generator $g$, and let  $x \in \mathbb{Z}$  is Alice’s private key, as well as  $y\in \mathbb{Z}$  is Bob’s private key. Alice publishes $g^x$ and Bob publishes $g^y.$ Then each of them computes the exchanged key $g^{xy} = (g^x)^y= (g^y)^x.$  
  
  In this paper, we apply and compare  two methods of algebraic cryptanalysis via linear algebra, namely, the linear decomposition method invented and developed by the author in \cite{R1} - \cite{R3} and in \cite{RM} (with A. Myasnikov), and the span-method invented and developed by B. Tsaban in \cite{T} and in \cite{TBK} (with A. Ben-Zvi and A. Kalka), to show vulnerability of the proposed in \cite{BKT} cryptosystem and protocol.  
  
  \section{The ElGamal-type cryptosystem   based on FACTOR \cite{BKT}}  
  
  Let $G$ be any public   group. Let $g, h \in G$ be two private elements of Alice,   and let $<g>$  and $<h>$ be the cyclic subgroups generated by these elements, respectively.   In order to define the FACTOR problem one assume that $< g > \cap  < h >= \{1\}$. Let $f: <g> \times <h> \rightarrow G$  be a function defined as follows: $f(g^x,h^y) = g^x \cdot h^y, $ where $x, y \in \mathbb{Z}.$  Obviously,  that $f$ is injective.  Then  FACTOR($g^xh^y$) $= f^{-1}(g^xh^y).$ 

{\bf Cryptosystem.} Let $G$ be a non-Abelian group and let $g,h \in G$ be two non commuting elements. We assume that $<g>\cap <h>=\{1\}.$ We  suppose that Alice is the recipient of the messages and Bob is communicating with Alice. Let $m \in  G$ be the message. 

Alice picks arbitrary  integers $x,y\in \mathbb{Z}$ and sets a public key  $(G,g,h,g^xh^y).$  

Alice has a private key $(g^x, h^y)$ for decryption.

To send the message $m$, Bob picks arbitrary   integers $x',y'$ and sends cipher text
$$ c = (g^{x+x'}h^{y+y'},g^{x'}h^{y'}m)$$
\noindent
 to Alice. 

To decrypt  the text, Alice uses her private key and  calculates 
$$(g^x)^{-1}(g^{x+x'}h^{y+y'})(h^y)^{-1}=g^{x'}h^{y'}.$$  
Then she  inverts it to retrieve $m.$ 

The authors of this scheme hoped  that the security of the crypto system described above reduces to solving FACTOR problem in the underlying group. Below we'll show that the system is vulnerable against linear algebra attacks. 

{\bf Cryptanalysis.}

We will show that any intruder can efficiently compute $g^{x'}h^{y'}$ and then retrieve $m.$ 

I. First we will use  the Tsaban's span-method. We suppose that $G$ is a finite group presented as a matrix group over a finite field. So, let 
 $G \leq$ M$_n$($\mathbb{F}_q$). Let $V=$ Lin$_{\mathbb{F}_q}(<g>)$ be the linear subspace of M$_n$($\mathbb{F}_q$) generated by all matrices of the form $g^i, i \in \mathbb{Z}.$ Then dim($V$) $\leq n-1.$ In fact, the matrices $1, g, g^2, ..., g^n$  are linearly dependent, since $g$ is the root of its characteristic polynomial of degree $n$. Obviously, if $g^{k+1}$ lies in Lin$_{\mathbb{F}_q}$($1, g, g^2, ..., g^k$), then $g^{k+t}, g^{1-t} \in$ Lin$_{\mathbb{F}_q}$($1, g^2, ..., g^k$) for every $t = 2, 3, ...$. 

We can efficiently construct a basis $1, g, g^2, ..., g^k$ of $V$ by checking for every succesive $l = 1, 2, ...$ either  $g^{l+1}$ lies in Lin$_{\mathbb{F}_q}$($1, g, g^2, ..., g^l$), or not. Then $k$ is the least $l$ such that this happens. Such verification is carried out by the Gauss elimination method which is known as efficient.  

Consider the equation 
\begin{equation}
\label{eq:1}
f(g^xh^y)h = hf(g^xh^y) \sim fg^xh = hfg^x,
\end{equation}
\noindent
that is linear with respect to $n^2$ unknown entries of matrix $f.$  We will seek $f$ in the form 
$$f = \sum_{i=0}^k\alpha_ig^i,$$
i.e., we  seek a solution $f$ in $V.$ We know that there is a non-degenerate solution $f = g^{-x}.$ 
We can efficiently construct a basis $e_1, ..., e_p$ of the subspace of all solutions of (\ref{eq:1}) in $V.$ Then we can use the following statement:

\medskip
{\bf Invertibility Lemma} \cite{T} (see also \cite{TBK}).

For a finite field  $\mathbb{F}_q$,  $e_1, ..., e_p \in $ M$_n$($\mathbb{F}_q$)), such that some linear combination of these matrices is invertible, if $\beta_1, ..., \beta_p$  are chosen uniformly and independently from $\mathbb{F}_q$, then the probability that the linear combination $f=\sum_{i=1}^p\beta_ie_i$ is invertible is at least $1- \frac{n}{q}$.

Let element $f$ be found. Then
$$f(g^xh^y) = h(fg^x), f(g^{x+x'}h^{y+y'})=(g^{x'}h^{y'})h^y(fg^x))$$
\noindent
and
$$(g^{x'}h^{y'})h^y(fg^x))f^{-1}(g^xh^y)^{-1}= g^{x'}h^{y'}.$$
So
$$(g^{x'}h^{y'})^{-1}(g^{x'}h^{y'}m)=m.$$ 
The message $m$ is recovered.

II. Now we will use the author's linear decomposition method. Let $G \leq$ M$_n$($\mathbb{F}$) be a matrix group over arbitrary (constructive) field $\mathbb{F}.$ Let $V=$ Lin$_{\mathbb{F}}(<g>(g^xh^y)<h>)$ be the linear subspace of M$_n$($\mathbb{F}$) generated by all matrices of the form $g^i(g^xh^y)h^j, i, j \in \mathbb{Z}.$ Then dim($V$) $\leq (n-1)^2.$ 

Let $e_1, e_2, ..., e_r$ be a basis of $V$ that can be efficiently obtained  in the same way as described above. Let $e_i = g^{u_i}(g^xh^y)h^{v_i}, u_i, v_i \in \mathbb{Z}, i = 1, ..., r.$ 

Since, $g^{x+x'}h^{y+y'} \in V,$ we can efficiently obtain a presentation of the form
\begin{equation}
\label{eq:2}
g^{x+x'}h^{y+y'} = \sum_{i=1}^r\alpha_ie_i, \ \alpha_i \in \mathbb{F}, \  i = 1, ..., r.
\end{equation}
The right side of (\ref{eq:2}) is equal to
\begin{equation}
\label{eq:3}
g^x(\sum_{i=1}^r\alpha_ig^{u_i}h^{v_i})h^y,
\end{equation}
\noindent
it follows by (\ref{eq:2}), that 
\begin{equation}
\label{eq:4}
g^{x'}h^{y'} = \sum_{i=1}^r\alpha_ig^{u_i}h^{v_i}.
\end{equation}
The message $m$ is recovered as above.

{\bf Remark.} Remind, that the authors of \cite{BKT} suggest as a platform for their cryptosystem  one of the groups GL$_n$($\mathbb{F}_q$), UT$_n$($\mathbb{F}_q$), or Braid Groups $B_n$, $n \in \mathbb{N}.$  In our cryptanalysis, we consider only matrix groups. Any group $B_n$ admits a faithful matrix representation \cite{Big}, \cite{Kramm}.    
 The braid group $B_n$  is linear via the so-called Lawrence-Krammer representation LK: $B_n \rightarrow$ GL$_m$($\mathbb{Z}[t^{\pm 1}, 1/2]$), where $m = n(n-1)/2,$
 is injective.
 The Lawrence–Krammer representation of a braid can be computed in polynomial time. This representation is also invertible in (similar) polynomial time (see  \cite{Kramm}, \cite{CJ}). 
  
\section{The Diffie-Hellman-type key exchange protocol   based on FACTOR \cite{BKT}} 
 
 Suppose Alice and Bob want to exchange keys. Suppose $G,g,h$ are as in FACTOR. Let Alice pick a pair of integers $(x_1,y_1),$ and Bob pick two integers $(x_2,y_2)$. 
 
 Then  Alice sends the element $g^{x_1}h^{y_1}$  to Bob. 
 
 Independently  Bob sends the element $g^{x_2}h^{y_2}$  to Alice.  
 
 Both Alice and Bob can recover the element $K=g^{x_1+x_2}h^{y_1+y_2}$. This is their private key. 

{\bf Cryptanalysis.}

Now we will apply and describe only the author's linear decomposition method. Let $G \leq$ M$_n$($\mathbb{F}$) be a matrix group over arbitrary (constructive) field $\mathbb{F}.$ Let $V=$ Lin$_{\mathbb{F}}(<g><h>)$ be the linear subspace of M$_n$($\mathbb{F}$) generated by all matrices of the form $g^ih^j, i, j \in \mathbb{Z}.$ Then dim($V$) $\leq (n-1)^2.$ 

Let $e_1, e_2, ..., e_r$ be a basis of $V$ that can be efficiently obtained  in the same way as described above. Let $e_i = g^{u_i}h^{v_i}, u_i, v_i \in \mathbb{Z}, i = 1, ..., r.$ 

Since, $g^{x_1}h^{y_1} \in V,$ we can efficiently obtain a presentation of the form
\begin{equation}
\label{eq:5}
g^{x_1}h^{y_1} = \sum_{i=1}^r\alpha_ie_i, \ \alpha_i \in \mathbb{F}, \  i = 1, ..., r.
\end{equation}
Then 
\begin{equation}
\label{eq:6}
\sum_{i=1}^r\alpha_ig^{u_i}(g^{x_2}h^{y_2})h^{v_i}= g^{x_2}( \sum_{i=1}^r\alpha_ie_i)h^{y_2}=K. 
\end{equation}
We succeeded again. 

Of course, the Tsaban's span-method can be applied too. 

  The described cryptanalysis has many analogues, presented in \cite{R1}-\cite{TBK}. In \cite{R4}, a general scheme based on multiplications is presented. It corresponds to a number of cryptographic systems known in the literature, which are also vulnerable to attacks by the linear decomposition method. Note that the Tsaban's span-method allows him to show the vulnerability of the well-known schemes of Anshel et al. \cite{AAG}, and   the Triple Decomposition Key Exchange Protocol of Peker \cite{Kurt}.
  
  A protection against linear algebra attacks is invented in \cite{R5}. It is described in the case of the Anshel et al. cryptographic scheme but can be applied to the Diffie-Hellman-type and some other schemes too.

\end{document}